\documentclass[twocolumn]{aastex6}

\def\lock102{HLock-102}
\def\helms34{HeLMS-34}

\received{24/04/2020}
\revised{18/05/2020}
\accepted{19/05/2020}
\usepackage{apjfonts}
\usepackage{aas_macros, xspace} 
\newcommand{\eq}{\,=\,}
\def\ts     {\thinspace}
\def\kms    {\ts km\ts s$^{-1}$}

\def\msol   {$M_{\odot}$}

\def\aco    {{\rm CO}($J$=1$\to$0)}

\def\bco    {{\rm CO}($J$=2$\to$1)}

\def\cco    {{\rm CO}($J$=3$\to$2)}
\def\dco    {{\rm CO}($J$=4$\to$3)}

\def\gco    {{\rm CO}($J$=7$\to$6)}

\def\jco    {{\rm CO}($J$=10$\to$9)}

\def\noema    {Northern Extended Millimeter Array (NOEMA)}
\shorttitle{VLASPECS:\ CO(1--0) in the HUDF}
\shortauthors{Riechers et al.}

\begin{document}

\title{${}$\vspace{-7.5mm} \\ VLA-ALMA Spectroscopic Survey in the Hubble Ultra Deep Field
  (VLASPECS): \\ Total Cold Gas Masses and CO Line Ratios for $z$=2--3
  ``Main Sequence'' Galaxies}

%% The new \altaffiliation can be used to indicate some secondary information
%% such as fellowships. This command produces a non-numeric footnote that is
%% set away from the numeric \affiliation footnotes.  NOTE that if an
%% \altaffiliation command is used it must come BEFORE the \affiliation call,
%% right after the \author command, in order to place the footnotes in
%% the proper location.
%%

\author{Dominik A.\ Riechers\altaffilmark{1,2}}
\author{Leindert A.\ Boogaard\altaffilmark{3}}
\author{Roberto Decarli\altaffilmark{4}}
\author{Jorge Gonz\'alez-L\'opez\altaffilmark{5}}
\author{Ian Smail\altaffilmark{6}}
\author{\\ Fabian Walter\altaffilmark{2,7}}
\author{Manuel Aravena\altaffilmark{8}}
\author{Christopher L.\ Carilli\altaffilmark{7,9}}
\author{Paulo C.\ Cortes\altaffilmark{10,11}}
\author{Pierre Cox\altaffilmark{12}}
\author{\\ Tanio D\'iaz-Santos\altaffilmark{8,13}}
\author{Jacqueline A.\ Hodge\altaffilmark{3}}
\author{Hanae Inami\altaffilmark{14}}
\author{Rob J.\ Ivison\altaffilmark{15}}
\author{Melanie Kaasinen\altaffilmark{2,16}}
\author{\\ Jeff Wagg\altaffilmark{17}}
\author{Axel Wei\ss\altaffilmark{18}}
\author{Paul van der Werf\altaffilmark{3}}

\altaffiltext{1}{Department of Astronomy, Cornell University, Space
  Sciences Building, Ithaca, NY 14853, USA}
\altaffiltext{2}{Max-Planck-Institut f\"ur Astronomie, K\"onigstuhl 17, D-69117 Heidelberg, Germany}
\altaffiltext{3}{Leiden Observatory, Leiden University, P.O.\ Box 9513, NL-2300 RA Leiden, The Netherlands}
\altaffiltext{4}{INAF-Osservatorio di Astrofisica e Scienza dello Spazio, via Gobetti 93/3, I-40129, Bologna, Italy}
\altaffiltext{5}{Carnegie Observatories, 813 Santa Barbara St, Pasadena, CA 91101, USA}
\altaffiltext{6}{Centre for Extragalactic Astronomy, Department of Physics, Durham University, South Road, Durham, DH1 3LE, UK}
\altaffiltext{7}{National Radio Astronomy Observatory, Pete V. Domenici Array Science Center, P.O.\ Box O, Socorro, NM 87801, USA}
\altaffiltext{8}{N\'ucleo de Astronom\'ia de la Facultad de Ingenier\'ia y Ciencias, Universidad Diego Portales, Av.\ Ej\'ercito Libertador 441, Santiago, Chile}
\altaffiltext{9}{Battcock Centre for Experimental Astrophysics, Cavendish Laboratory, Cambridge CB3 0HE, UK}
\altaffiltext{10}{Joint ALMA Observatory - ESO, Av.\ Alonso de C\'ordova, 3104, Santiago, Chile}
\altaffiltext{11}{National Radio Astronomy Observatory, 520 Edgemont Rd, Charlottesville, VA, 22903, USA}
\altaffiltext{12}{Sorbonne Universit{\'e}, UPMC Universit{\'e} Paris 6 and CNRS, UMR 7095, Institut d'Astrophysique de Paris, 98bis boulevard Arago, F-75014 Paris, France}
\altaffiltext{13}{Chinese Academy of Sciences South America Center for Astronomy (CASSACA), National Astronomical Observatories, CAS, Beijing 100101, China}
\altaffiltext{14}{Hiroshima Astrophysical Science Center, Hiroshima University, 1-3-1 Kagamiyama, Higashi-Hiroshima, Hiroshima 739-8526, Japan}
\altaffiltext{15}{European Southern Observatory, Karl-Schwarzschild-Stra{\ss}e 2, D-85748 Garching, Germany}
\altaffiltext{16}{Universit\"{a}t Heidelberg, Zentrum f\"{u}r Astronomie, Institut f\"{u}r Theoretische Astrophysik, Albert-Ueberle-Stra\ss e 2, D-69120 Heidelberg, Germany}
\altaffiltext{17}{SKA Organization, Lower Withington Macclesfield, Cheshire SK11 9DL, UK}
\altaffiltext{18}{Max-Planck-Institut f\"ur Radioastronomie, Auf dem H\"ugel 69, D-53121 Bonn, Germany}
 \email{riechers@cornell.edu}

\begin{abstract}

Using the NSF's Karl G.\ Jansky Very Large Array (VLA), we report six
detections of \aco\ emission and one upper limit in $z$=2--3 galaxies
originally detected in higher-$J$ CO emission in the Atacama
  Large submillimeter/Millimeter Array (ALMA) Spectroscopic Survey in
the Hubble Ultra Deep Field (ASPECS). From the \aco\ line strengths,
we measure total cold molecular gas masses of $M_{\rm
  gas}$=2.4--11.6$\times$10$^{10}$ ($\alpha_{\rm CO}$/3.6)\,\msol.
We also measure a median
\cco\ to \aco\
line brightness temperature ratio of
$r_{31}$=0.84$\pm$0.26, and a \gco\ to \aco\ ratio range of
$r_{71}$$<$0.05 to 0.17.
These results suggest that \cco\ selected galaxies may have a higher
CO line excitation on average than \aco\ selected galaxies, based on
the limited, currently available samples from the ASPECS and VLA CO
Luminosity Density at High Redshift (COLDz) surveys. This implies that
previous estimates of the cosmic density of cold gas in galaxies based
on \cco\ measurements should be revised down by a factor of $\simeq$2
on average based on assumptions regarding CO excitation alone. This
correction further improves the agreement between the best currently
existing constraints on the cold gas density evolution across cosmic
history from line scan surveys, and the implied characteristic gas
depletion times.

\end{abstract}


\section{Introduction} \label{sec:intro}

Detailed studies of the star formation history of the universe, i.e.,
the volume density of star formation activity with redshift, have
shown that, $\sim$10\,billion years ago, ``typical'' and starburst
galaxies were forming 10--30 times more stars per year than at the
present day. The observed buildup of stars is consistent with
measurements of the volume density of stellar mass in galaxies through
cosmic times (see, e.g., Madau \& Dickinson \citeyear{md14} for a
review). Studies of the cold molecular gas, the prospective fuel for
star formation, and gas mass fractions in high redshift galaxies (see,
e.g., Carilli \& Walter \citeyear{cw13}; Combes \citeyear{combes18}
for reviews), suggest that this higher star formation activity is
primarily due to an increased availability of fuel, rather than
fundamental differences in the star formation process at earlier
epochs (e.g., Daddi et al.\ \citeyear{daddi10a}; Riechers et
al.\ \citeyear{riechers11e}; Ivison et al.\ \citeyear{ivison11};
Tacconi et al.\ \citeyear{tacconi13,tacconi18}; Bothwell et
al.\ \citeyear{bothwell13}; Genzel et al.\ \citeyear{genzel15};
Scoville et al.\ \citeyear{scoville17}; Kaasinen et
al.\ \citeyear{kaasinen19}).

The rise of a new generation of powerful radio to sub/millimeter
wavelength interferometers such as the NSF's Karl G.\ Jansky Very
Large Array (VLA), the Atacama Large submillimeter/Millimeter Array
(ALMA), and \noema\ over the past decade is now enabling the first
comprehensive view of the baryon cycle, i.e., the conversion from gas
to stars over cosmic time, unveiling how galaxies grow across the
history of the universe.
This has only recently become possible based on the first large cosmic
volume surveys for the cold gas density evolution at high redshift
through the VLA CO Luminosity Density at High Redshift (COLDz; e.g.,
Pavesi et al.\ \citeyear{pavesi18b}; Riechers et
al.\ \citeyear{riechers19a}) and ALMA Spectroscopic Survey in the
Hubble Ultra Deep Field (ASPECS; e.g., Walter et
al.\ \citeyear{walter16}; Decarli et al.\ \citeyear{decarli19}) CO
line scan surveys. Together with an earlier pilot study with
PdBI/NOEMA in the Hubble Deep Field (Decarli et
al.\ \citeyear{decarli14}; Walter et al.\ \citeyear{walter14}), these
surveys have now covered a volume approaching 500,000\,Mpc$^3$.
  In the most sensitive areas, these studies reach down to galaxies
  below the characteristic CO luminosity $L_{\rm CO}^{\star}$ out to
  at least $z$$\sim$3, showing that they select representative
  star-forming galaxies at high redshift. Despite the fact that they
cover different survey fields, the cosmic gas density measurements of
ASPECS and COLDz are remarkably consistent, showing that cosmic
variance likely is not the dominant source of uncertainty of the
measurements at this stage (see also Popping et
al.\ \citeyear{popping19}). However, one remaining source of
uncertainty is due to the fact that these surveys cover different CO
transitions in the overlapping redshift ranges. In particular, COLDz
measures \aco\ emission at $z$=2--3, while ASPECS measures
\cco\ emission at the same redshift. To address possible uncertainties
due to CO excitation, the ASPECS measurements are ``corrected'' by
adopting a
\cco\ to \aco\ line brightness temperature ratio of
$r_{31}$=0.42$\pm$0.07, based on previous measurements of three ``main
sequence'' galaxies at $z$$\simeq$1.5 (i.e., the closest comparison
sample available at the time; Daddi et al.\ \citeyear{daddi15}) before
adopting an $\alpha_{\rm CO}$ conversion factor to translate the
inferred \aco\ line luminosities to gas masses (see Decarli et
al.\ \citeyear{decarli19} for details).

We here present VLA observations of the Hubble Ultra Deep Field (HUDF)
in a region covered by ASPECS at higher frequencies (i.e., in
higher-$J$ CO lines) to derive more robust estimates of CO line
brightness temperature ratios for gas-selected galaxies
by constraining the gas excitation in the low-$J$ CO lines. We use
these data to measure total cold molecular gas masses, gas depletion
times, and baryonic gas mass fractions. Our observations cover seven
of the eight ASPECS sources in the $z$=2--3 redshift range, and thus
provide direct measurements of most of the sources that are used to
infer the cosmic gas density measurements near the peak of the cosmic
star formation history in this field. We describe the observations in
Sect.\ 2, and present the results in Sect.\ 3. Further analysis and a
discussion of the impact of our findings are given in Sect.\ 4, before
we provide a summary and conclusions in Sect.\ 5. We use a
concordance, flat $\Lambda$CDM cosmology throughout, with
$H_0$\eq69.6\,\kms\,Mpc$^{-1}$, $\Omega_{\rm M}$\eq0.286, and
$\Omega_{\Lambda}$\eq0.714 (Bennett et al.\ \citeyear{bennett14}).

\section{Data} \label{sec:data}

\begin{figure*}[tbh]
\epsscale{1.15}
\plotone{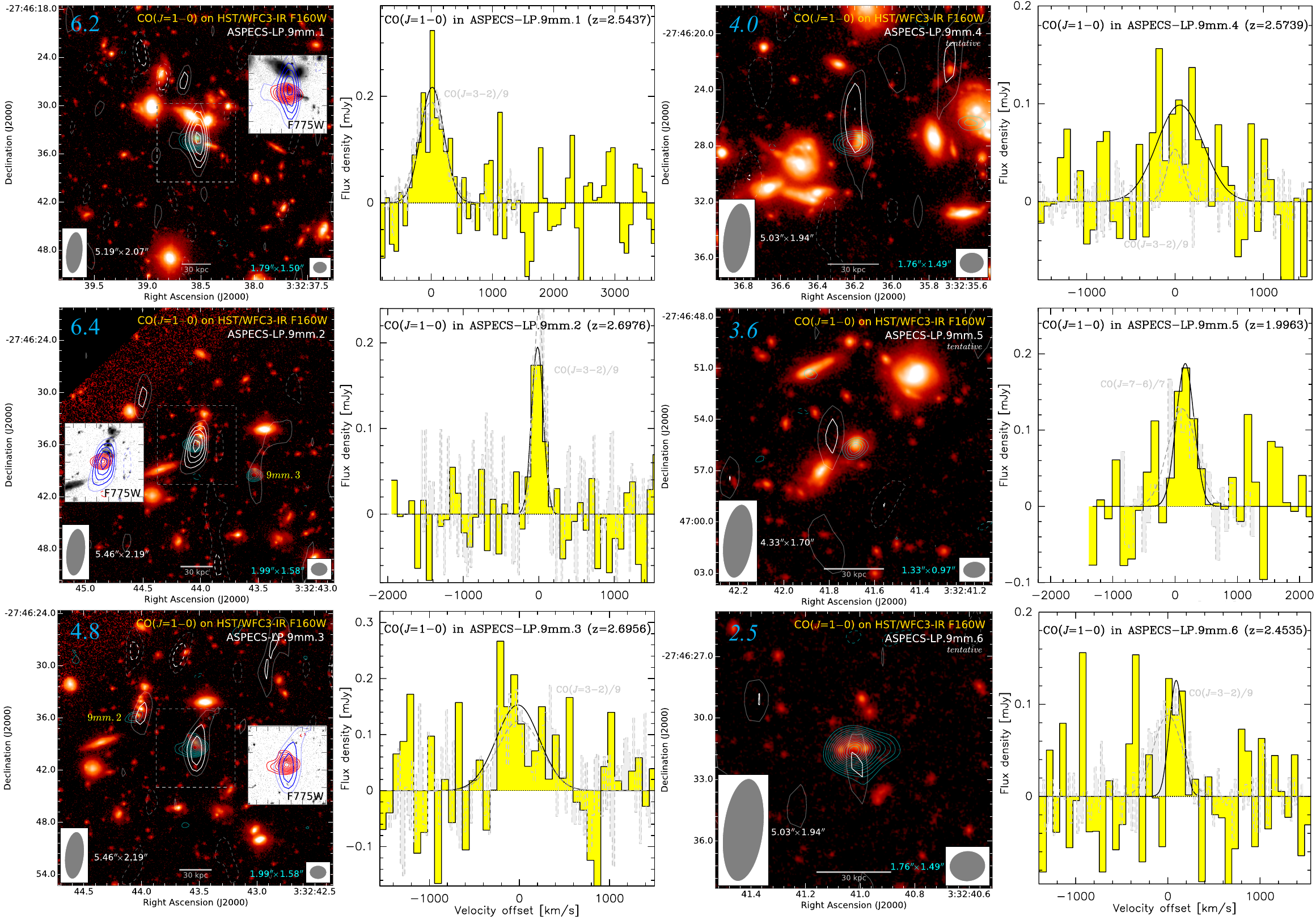}
\vspace{-2.0mm}

\caption{VLA \aco\ moment-0 line maps ({\em left panels;} white
  contours) and line spectra ({\em right panels;} solid histograms) of
  all detected galaxies in the sample, and Gaussian fits to the line
  profiles (black curves) where applicable. Contour maps are shown
  overlaid on HST/WFC3-IR F160W images, and ACS F775W insets for the
  robust detections (Illingworth et al.\ \citeyear{illingworth13}),
  with \aco\ peak signal-to-noise ratios indicated in blue in
  the top left corner of
  each map panel. ALMA \cco\ or \gco\ (9mm.5 only) contours from
  Gonzalez-Lopez et al.\ (\citeyear{gl19}) or B20 are shown for
  comparison (aqua color). VLA maps are integrated over
  737, 192, 923, 632, 405, and 341\,\kms\
    (80, 20, 96, 68, 52, and 38\,MHz),
    for 9mm.1, 2, 3, 4, 5, and 6, respectively. VLA contour levels
  are in steps of
  1$\sigma$=14, 25, 12.5, 14, 30, and 20.2\,$\mu$Jy\,beam$^{-1}$,
  starting at $\pm$2$\sigma$ (except 9mm.6, where an additional
  2.5$\sigma$ contour level is shown). ALMA contour levels are in
  steps of
  1$\sigma$=27, 65, 38, 37, 105, and 29\,$\mu$Jy\,beam$^{-1}$,  
  starting at $\pm$3$\sigma$, except 9mm.1,
  where contour steps are $\pm$3$\sigma$. The VLA (ALMA) beam sizes
  are shown in the bottom left (right) corner of each panel. 9mm.5
  shows an offset between the peak position of both lines, likely
  primarily due to the modest signal-to-noise ratio of the tentative
  \aco\ detection. Spectra are shown at resolutions of
  74, 77, 77, 74, 149, and 72\,\kms\
  (8, 8, 8, 8, 16, and 8\,MHz),
  respectively. Velocity scales are relative to the redshifts
  indicated.  Scaled ALMA \cco\ or \gco\ spectra (dashed gray
  histograms; Gonzalez-Lopez et al.\ \citeyear{gl19}; B20) are shown
  for comparison.
    \label{f1}}
%\vspace{-5mm}
%
\end{figure*}

We used the VLA to observe redshifted \aco\ emission (rest-frame
frequency:\ $\nu_{\rm rest}$=115.2712\,GHz) in seven galaxies in the
HUDF at $z$=2.0--2.7 (VLA program ID:\ 19B-131; PI:\ Riechers). We
used the Ka band receivers in combination with the WIDAR correlator
configured to 3-bit sampling mode to observe a contiguous bandwith of
8\,GHz (dual polarization) covering the 30.593--38.662\,GHz (i.e.,
$\sim$9\,mm) frequency range at 2\,MHz spectral resolution
(17\,\kms\ at 35\,GHz). Some minor overlaps between subbands were
employed to avoid that the centers of known lines fall onto subband
gaps. Gaps between subbands were mitigated by employing three
frequency switching setups, shifted by $\pm$12\,MHz relative to the
central setup. To cover all targets, as well as $\sim$120 fainter
galaxies with secure optical spectroscopic redshifts for which the
\aco\ or \bco\ line is accessible within our data set, two telescope
pointings centered at J2000 03:32:43.294, $-$27:46:44.88 and
03:32:38.834 $-$27:46:35.46 were observed to equal depth.
Observations were carried out under very good weather conditions in D
array using 17\,scheduling blocks with a length of 2.5\,hr each
between 2019 December 07 and 2020 January 27. This resulted in a total
time of 42.5\,hr, or 14.7\,hr on source per pointing.\footnote{A total
  of 82.5\,hr were approved, but could not be completed due to weather
  and scheduling constraints given the low declination of the HUDF.}
Given the declination of the HUDF, four of the 27 antennas were
shadowed by other antennas and thus flagged in all data sets. The
radio quasar J0348$-$2749 ($S_\nu$=1.79$\pm$0.13\,Jy based on our
  calibration, which provides individual values covering the
  1.61--1.99\,Jy range) was observed every 9\,minutes for complex
gain calibration. The quasar 3C\,48 ($S_\nu$=0.70 to 0.88\,Jy
  from the upper to the lower frequency edges of the bandpass, based
  on the Perley \& Butler \citeyear{pb17} scale) was observed once
per scheduling block for flux calibration. Given its recent flaring
activity,\footnote{See {\tt
    https://science.nrao.edu/facilities/vla/docs/
    manuals/oss/performance/fdscale}, version 2019 November 19.} we
conservatively consider the absolute flux calibration to be reliable
at the $\sim$15\% uncertainty level.

All data were processed with the CASA 5.6.2 pipeline, augmented by
manual data editing where necessary. Imaging the data in mosaicking
mode with natural baseline weighting out to the 10\% primary beam
response\footnote{The VLA primary beam full width at half power at our
  observing frequencies is $\sim$65$''$--82$''$.} region yields a
synthesized clean beam size of 4.99$''$$\times$1.96$''$ (largest
recoverable scale:\ $\sim$45$''$) and an rms noise level of
1.8\,$\mu$Jy\,beam$^{-1}$ across the entire 8\,GHz continuum bandwidth
covered by the spectral setup. The noise level increases by nearly a
factor of two from the low- to the high-frequency edge of the
bandpass, as expected based on the increasing receiver and atmospheric
noise temperatures with frequency in the Ka band. The rms noise in the
phase centers is 40--44\,$\mu$Jy\,beam$^{-1}$ per 75\,\kms\ bin at the
line frequencies of all targets except the lowest-redshift source,
where it is 70\,$\mu$Jy\,beam$^{-1}$.

%%%%%%%%%%%%%%%%%%%%%%%%%%%%%%%%%%%%%%%%%%%%%%%%%
%%%% Tab.1 Line Parameters
%%%%%%%%%%%%%%%%%%%%%%%%%%%%%%%%%%%%%%%%%%%%%%%%%

\begin{figure*}[tbh]
\begin{deluxetable}{ l l c c c c c c c c c c c }
\tabletypesize{\scriptsize}
\tablecaption{VLASPECS line parameters. \label{t1}}
\tablehead{
 VLA ID & ALMA ID & $z_{\rm ALMA}$ & $I_{\rm CO(1-0)}$ & dv$_{\rm CO(1-0)}$ & dv$_{\rm ALMA}$$^\tablenotemark{a}$ & $L'_{\rm CO(1-0)}$ & $M_{\rm gas}$ &  $f_{\rm gas}$$^\tablenotemark{b}$ & $f_{\rm bary}$$^\tablenotemark{c}$ & $t_{\rm dep}$$^\tablenotemark{d}$ & $r_{31}$ & $r_{71}$ \\
        &         &     & [Jy\,\kms]      & [\kms ]         & [\kms ] & [10$^{10}$\,K\,\kms\,pc$^2$] & [10$^{11}$\,\msol] & & & [Gyr] & }
\startdata
9mm.1   & 3mm.1    & 2.5437 & 0.103$\pm$0.022 & 447$\pm$110 & 519$\pm$18  & 3.22$\pm$0.68 & 1.16$\pm$0.24 & 4.6 & 0.82 & 0.46 & 0.84$\pm$0.18 & 0.17$\pm$0.05 \\
9mm.2   & 3mm.9    & 2.6976 & 0.038$\pm$0.006 & 201$\pm$47 & 166$\pm$24  & 1.32$\pm$0.19 & 0.48$\pm$0.07 & 0.38 & 0.27 & 0.15 & 1.10$\pm$0.21 & $<$0.93 \\
9mm.3   & 3mm.7    & 2.6956 & 0.091$\pm$0.022 & 560$\pm$230 & 570$\pm$70  & 3.14$\pm$0.75 & 1.13$\pm$0.27 & 0.90 & 0.47 & 0.57 & 0.79$\pm$0.21 & $<$0.21 \\
{\em 9mm.4}$^\star$ & 3mm.12   & 2.5739 & {\em 0.064$\pm$0.019} & {\em 620$\pm$280} & 221$\pm$40  & {\em 2.03$\pm$0.60} & {\em 0.73$\pm$0.22} & {\em 1.84} & {\em 0.65} & {\em 2.3} & {\em 0.23$\pm$0.08} & {\em $<$0.05} \\
{\em 9mm.5}$^\star$ & 1mm.C14a & 1.9963 & {\em 0.065$\pm$0.018} & {\em 342$\pm$96} & 281$\pm$57  & {\em 1.31$\pm$0.37} & {\em 0.47$\pm$0.13} & {\em 0.75} & {\em 0.43} & {\em 0.94} & --- & {\em 0.12$\pm$0.04} \\
{\em 9mm.6}$^\star$ & 3mm.3    & 2.4535 & {\em 0.022$\pm$0.009} & {\em 176$\pm$110} & 367$\pm$31  & {\em 0.66$\pm$0.26} & {\em 0.24$\pm$0.09} & {\em 0.47} & {\em 0.32} & {\em 0.37} & {\em 1.54$\pm$0.61} & {\em $<$0.25} \\
9mm.7 & 1mm.C07  & 2.5805 & $<$0.105 (3$\sigma$) & --- & 660$\pm$110 & $<$3.4 & $<$1.2 & $<$1.2 & $<$0.55 & $<$3.0 & $>$0.17 & $>$0.09 \\
%\vspace{-2mm}
\enddata
\tablecomments{Stellar masses, star formation rates, and $J_{\rm upper}$$\geq$3 CO line parameters used in the calculations were adopted from Gonzalez-Lopez et al.\ (\citeyear{gl19}) and B20 (see also Aravena et al.\ \citeyear{aravena19}; M.\ Aravena et al.\ 2020, in prep.). VLA primary beam correction factors of
  pbc=0.984, 0.913, 0.970, 0.565, 0.785, 0.894, and 0.286
  were adopted throughout for 9mm.1, 2, 3, 4, 5, 6, and 7, respectively. We here report \aco\ line parameters based on a signal-to-noise ratio optimized extraction, i.e., without tying them to the ALMA measurements. Fixing the extraction to the ALMA-based line centroids and widths would yield changes in $r_{31}$ by
  6.4\%, --11\%, 0.6\%, --57\%, and 3.6\%
  for 9mm.1, 2, 3, 4, and 6, respectively, or --0.08\% on average when excluding 9mm.4. These differences are negligible compared to other sources of uncertainty for all sources except 9mm.4. Where not provided, we assume uncertainties of 25\% for robustly \aco-detected sources, and 40\% for tentatively-detected sources. $r_{31}$ and $r_{71}$ are \cco\ to \aco\ and \gco\ to \aco\ line brightness temperature ratios, respectively.}
\tablenotetext{\star}{Tentative detection; independent confirmation of line parameters from more sensitive data required.}
\tablenotetext{\rm a}{Obtained from a simultaneous fit of all ALMA-detected CO/[CI] lines considered by B20, i.e., excluding the VLA \aco\ measurements reported here.}
\tablenotetext{\rm b}{Defined as $f_{\rm gas}$=$M_{\rm gas}$/$M_\star$; also commonly referred to as the gas-to-stellar mass ratio $\mu_{\rm mol}$ or $\mu_{\rm gas}$ in the literature.}
\tablenotetext{\rm c}{Defined as $f_{\rm bary}$=$M_{\rm gas}$/($M_{\rm gas}$+$M_\star$).}
\tablenotetext{\rm d}{Defined as $t_{\rm dep}$=$M_{\rm gas}$/SFR.}
\end{deluxetable}
\end{figure*}

%\vspace{-50mm}

%%%%%%%%%%%%%%%%%%%%%%%%%%%%%%%%%%%%%%%%%%%%%%%%%%%%%%%%%%%

\section{Results} \label{sec:results}

We robustly detected \aco\ emission towards three targets at
$>$4.5$\sigma$ significance and tentatively detect another three at
$\sim$2.5--4.0$\sigma$, but we have not detected the seventh target
(which lies in a region where sensitivity is reduced by a factor of
$\sim$3.5 due to primary beam attenuation).
We extracted spectra at the map peak positions of all targets from the
primary beam corrected mosaic (or from the ALMA CO $J$=3$\to$2 peak
position for the non-detection), and fitted them with Gaussian line
profiles (Fig.~\ref{f1} and Appendix). We then created moment-0
maps across the velocity ranges where emission is seen in the spectra.
ASPECS-LP.9mm.1, 2, 3, 4, 5, and 6 (hereafter:\ 9mm.1 to 6)
are detected at peak signal-to-noise ratios of 6.2, 6.4, 4.8, 4.0,
3.6, and 2.5,
respectively, in the moment-0 maps (Fig.~\ref{f1}).
We then fitted two-dimensional Gaussian profiles in the image plane to
the emission in the moment-0 maps to investigate if sources are
extended.\footnote{Uncertainties from these fits are propagated to the
  reported line fluxes.} All sources except 9mm.1 and 3 are consistent
with point sources. 9mm.1 has a formal deconvolved size of
(4.7$\pm$2.4)$\times$(1.1$\pm$0.7)\,arcsec$^2$, which corresponds to
(39$\pm$19)$\times$(9$\pm$6)\,kpc$^2$. 9mm.3 has a formal deconvolved
size of (4.7$\pm$2.0)$\times$(1.0$\pm$1.6)\,arcsec$^2$, which
corresponds to (38$\pm$16)$\times$(8$\pm$13)\,kpc$^2$.\footnote{9mm.5
  is best fitted with a finite size, resulting in a formal deconvolved
  size of (1.7$\pm$1.8)$\times$(0.4$\pm$1.5)\,arcsec$^2$, which
  corresponds to (14$\pm$15)$\times$(3$\pm$13)\,kpc$^2$. Given that
  the source is only tentatively detected, and the resulting
  significant uncertainties, we only consider this a weak constraint.}
Both sources are smaller than the beam, and thus, are marginally
resolved along their source major axes (which are close to the VLA
beam minor axes) at best. The extension of 9mm.1 appears to be
consistent with that seen in the ALMA \cco\ data
(Fig.~\ref{f1}). Future observations at higher resolution and greater
sensitivity are necessary to better constrain the true sizes of these
galaxies.

All line fluxes and widths and the corresponding line luminosities are
summarized in Table~\ref{t1}. The line widths agree with those
measured from the higher-$J$ lines observed by ALMA within the
uncertainties, with two exceptions. The tentatively-detected
\aco\ line in 9mm.6 is narrower than the \cco\ line by a factor of
2.1$\pm$1.4. Upon inspection of the line profile, it becomes clear
that the limited signal-to-noise ratio of the measurement did not
allow for a detection of the faint blue line component seen by ALMA
(which also causes a small offset in the peak velocities), such that
the \aco\ line width may be biased low because the Gaussian fit does
not account for this component. On the other hand, the
tentatively-detected \aco\ line in 9mm.4 is 2.8$\pm$1.4 times broader
than its \cco\ line. Although the uncertainties are still significant,
this may indicate the presence of an extended cold gas reservoir with
low gas excitation, which could be partially missed in the higher-$J$
CO line measurements.
No higher-$J$ lines than \cco\ are detected in this source by ALMA
(see L.\ Boogaard et al.\ 2020, in prep.\ [B20 hereafter], for further
details). However, this finding needs to be investigated further with
more sensitive data. 9mm.5 may show slight differences in the line
profiles between \aco\ and \gco, which is consistent with a minor
difference in peak velocities.
If confirmed, this could be due to differential gas excitation across
the galaxy, but no firm conclusions are possible at the current
signal-to-noise ratio of the data (which likely is also responsible
for the apparent spatial offset between the emission peaks). In all
other cases, the peak position of the \aco\ emission coincides with
that of the higher-$J$ CO emission and the stellar light within the
uncertainties (Fig.~\ref{f1}).

We convert the \aco\ line luminosities to total cold molecular gas
masses by adopting a conversion factor of $\alpha_{\rm
  CO}$=3.6\,\msol\,(K\,\kms\,pc$^2$)$^{-1}$, as was done in our
previous work (e.g., Riechers et al.\ \citeyear{riechers19a}; Decarli
et al.\ \citeyear{decarli19}; see also Daddi et
al.\ \citeyear{daddi10a}), in consistency with the stellar
mass--metallicity relation (Boogaard et al.\ \citeyear{boogaard19};
see also Aravena et al.\ \citeyear{aravena19}).\footnote{Since
    the calibration of $\alpha_{\rm CO}$ depends on the ratio of the
    gas density $n$ and the CO line brightness temperature $T_{\rm b}$
    ($\alpha_{\rm CO}$$\propto$$\sqrt{n}$\,$T_{\rm b}$$^{-1}$ in the
    simplest case; e.g., Solomon \& Vanden Bout \citeyear{sv05};
    Bolatto et al.\ \citeyear{bolatto13}), it is expected to scale
    with CO excitation in practice. Our current constraints for the
    ASPECS sample appear to disfavor significantly lower $\alpha_{\rm
      CO}$ values than adopted in this work, but dynamical mass
  measurements from higher-resolution CO observations in the future
  will be required to more directly calibrate $\alpha_{\rm CO}$.} We
also measure line brightness temperature ratios relative to the
\cco\ and \gco\ lines, using the line fluxes measured from the ALMA
data (B20, and references therein).

\begin{figure*}
\epsscale{1.15}
\plotone{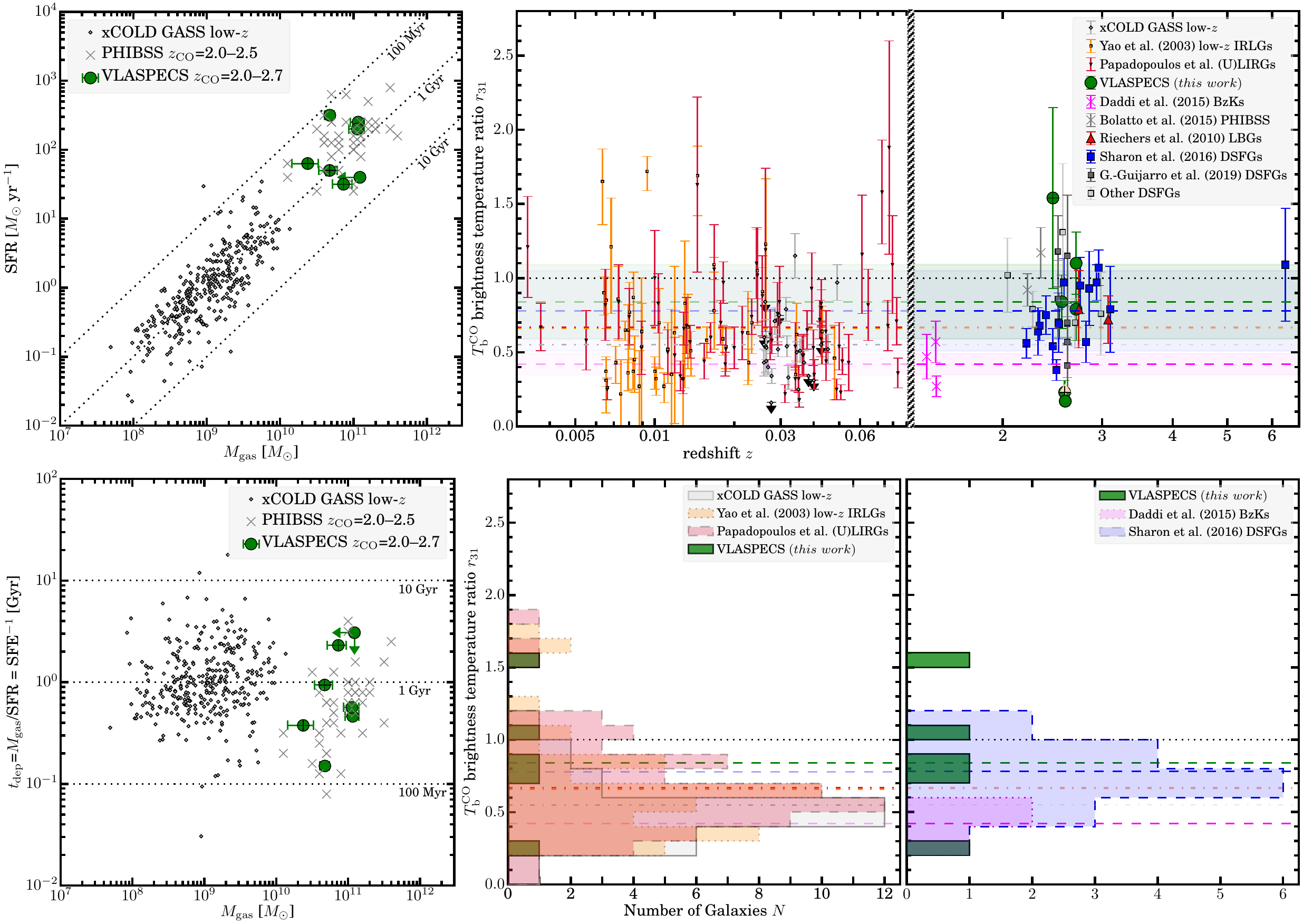}
\vspace{-1.5mm}

\caption{{\em Top left:} The revised, \aco-based $M_{\rm gas}$ from
  VLASPECS confirm that $z$=2--3 galaxies detected in the ASPECS
  survey (green circles; tentative detections are marked with a plus sign)
  closely follow the ``star formation law''
  (i.e., $M_{\rm gas}$--SFR relation) at high redshift. CO-detected
  ``main sequence'' galaxies at similar redshifts from the PHIBBS1/2
  surveys (typically based on CO $J$=3$\to$2, but using a
  metallicity-dependent conversion factor; Tacconi et
  al.\ \citeyear{tacconi18}) and local galaxies from the xCOLD GASS
  \aco\ survey (Saintonge et al.\ \citeyear{saintonge17}) are shown
  for comparison. {\em Bottom left:} Same, but plotting the depletion
  time $t_{\rm dep}$ against $M_{\rm gas}$. All samples cover a
  similar range in $t_{\rm dep}$, but the average $t_{\rm dep}$ for
  the (higher $M_{\rm gas}$) high-$z$ samples appear lower. {\em Top
    right:} The $r_{31}$ brightness temperature ratio of VLASPECS
  galaxies (green circles) is similar to that of strongly-lensed
  $z$$\sim$3 Lyman-break galaxies (red triangles; Riechers et
  al.\ \citeyear{riechers10c}), $z$$>$2 ``main sequence'' galaxies
  from the PHIBSS survey (gray crosses; Bolatto et
  al.\ \citeyear{bolatto15}), and $z$$>$2 dusty star-forming galaxies
  (DSFGs; blue squares; compilation from Sharon et
  al.\ \citeyear{sharon16}, including data from Riechers et
  al.\ \citeyear{riechers11c,riechers11d,riechers13b}; Ivison et
  al.\ \citeyear{ivison11}; Danielson et al.\ \citeyear{danielson11};
  Thomson et al.\ \citeyear{thomson12}; Fu et al.\ \citeyear{fu13};
  Sharon et al.\ \citeyear{sharon13,sharon15}; other DSFGs shown as
  light gray squares are from Nayyeri et al.\ \citeyear{nayyeri17};
  Dannerbauer et al.\ \citeyear{dannerbauer19}; Harrington et
  al.\ \citeyear{harrington19}; Leung et al.\ \citeyear{leung19};
  Sharon et al.\ \citeyear{sharon19}) and clustered DSFGs (dark gray
  squares; Bussmann et al.\ \citeyear{bussmann15}; Gomez-Guijarro et
  al.\ \citeyear{gomez19}), but $\sim$2 times higher on average than
  BzK-selected ``main sequence'' galaxies at $z$$\sim$1.5 (magenta
  crosses; Daddi et al.\ \citeyear{daddi15}). Nearby galaxy samples
  from the xCOLD GASS survey (Lamperti et al.\ \citeyear{lamperti20})
  and two studies of infrared-luminous galaxies (Yao et
  al.\ \citeyear{yao03}; Papadopoulos et
  al.\ \citeyear{papadopoulos12}) are shown for comparison.  Dashed
  lines and shaded regions indicate mean/median values and spread for
  high-$z$ samples with $>$2 galaxies or clusters, with the same color
  coding as the symbols. Dash-dotted lines indicate mean values for
  the low-$z$ samples. {\em Bottom right:} Same, but shown as binned
  histograms in $r_{31}$ (excluding upper limits) and across the full
  redshift range, and only including samples for which mean/median
  values are indicated in the {\em top right} panels.
  \label{f4}}
%\vspace{-5mm}
%
\end{figure*}

\section{Analysis and Discussion} \label{sec:analysis}

\subsection{Gas masses, depletion times, and line ratios}

We find total cold molecular gas masses of
4.8--11.6$\times$10$^{10}$\,\msol\ for our sample
(2.4--11.6$\times$10$^{10}$\,\msol\ when including tentative
detections), which corresponds to baryonic gas mass fractions of
27\%--82\%, and gas depletion times of 150--570\,Myr
(150\,Myr--2.3\,Gyr when including tentative detections; see
Table~\ref{t1}). These galaxies thus follow the ``star formation law''
(i.e., $M_{\rm gas}$--SFR relation) for ``main sequence'' galaxies at
high redshift (Fig.~\ref{f4} {\em left}). Only 9mm.2 shows a short gas
depletion time, as is characteristic of starburst
galaxies.\footnote{Gas depletion times depend on the conversion
  factor, and would be shorter for lower $\alpha_{\rm CO}$ in
  principle.}

We measure CO line brightness temperature ratios between the \cco\ and
\aco\ lines of $r_{31}$=0.79--1.10 for the robust line detections, or
0.23--1.54 when including tentative detections,
with a median value of 0.84$\pm$0.05 or 0.84$\pm$0.26
and a mean value of 0.91$\pm$0.14 or 0.90$\pm$0.43 when excluding or
including tentative detections, respectively.\footnote{Quoted
  uncertainties are one standard deviation for the mean and the median
  absolute deviation for the median, and exclude absolute flux
  calibration uncertainties between the VLA and ALMA observations.}
This is comparable to the mean line ratios found for strongly-lensed
Lyman-break galaxies (Fig.~\ref{f4} {\em right}; $\sim$0.75; Riechers
et al.\ \citeyear{riechers10c}) and dusty star-forming galaxies at
similar redshifts (0.78$\pm$0.27; Sharon et al.\ \citeyear{sharon16};
see also, e.g., Riechers et al.\ \citeyear{riechers11c,riechers11d};
Ivison et al.\ \citeyear{ivison11}; Danielson et
al.\ \citeyear{danielson11}; Thomson et al.\ \citeyear{thomson12};
Frayer et al.\ \citeyear{frayer18}), but twice as high as the value of
$r_{31}$=0.42$\pm$0.07 adopted in previous works (based on a sample of
three $z$$\sim$1.5 ``main sequence'' galaxies from Daddi et
al.\ \citeyear{daddi15}), suggesting that the gas masses at
$z$$\sim$2.5 estimated based on the ALMA measurements of the
  \cco\ line alone should be corrected down by a factor of $\simeq$2
on average.

We also find line brightness temperature ratios between the \gco\ and
\aco\ lines of $r_{71}$$<$0.05--0.17, with additional, less
constraining upper limits in the $<$0.21 to $<$0.93 range. The only
robust detection in both lines is 9mm.1, with
$r_{71}$=0.17$\pm$0.05. Our findings suggest that, in lieu of
observational constraints, $r_{71}$=0.1--0.2 may be considered a
reasonable assumption for $z$=2--3 ``main sequence'' galaxies, 
  but we caution that 9mm.1 contains an active galactic nucleus (AGN).\footnote{
  The galaxies 9mm.2 and 4 in our sample also contain AGN
  (Luo et al.\ \citeyear{luo17};
  Boogaard et al.\ \citeyear{boogaard19}).}
 This is comparable to the characteristic value proposed for dusty
 star-forming galaxies at similar redshifts ($r_{71}$=0.18$\pm$0.04;
 Bothwell et al.\ \citeyear{bothwell13}). It is also comparable
   to the mean value found for a sample of nearby luminous and
   ultra-luminous infrared galaxies studied by Rosenberg et
   al.\ (\citeyear{rosenberg15}), i.e., $r_{71}$=0.15$\pm$0.10, but
   below the most highly-excited sources in that sample (their ``Class
   III'' objects), $r_{71}$=0.24$\pm$0.11. The latter subsample
   includes those galaxies for which an AGN contribution to the line
   excitation is the most plausible (such as Mrk\,231; e.g., van der
   Werf et al.\ \citeyear{vdw10}), but it should be noted that current
   evidence indicating that AGN lead to changes in $r_{71}$ remains
   ambiguous at best.\footnote{We also caution that CO line ratios at
     high redshift are impacted by the warmer cosmic microwave
     background (CMB), which could increase $r_{71}$ in the presence
     of low excitation, low brightness temperature gas (e.g., da Cunha
     et al.\ \citeyear{dacunha13b}).} As an example, in the CO line
   excitation model for Mrk\,231 shown by van der Werf et
   al.\ (\citeyear{vdw10}), the starburst contribution to the
   \gco\ flux is about three times higher than that by the
   AGN. Moreover, Lu et al.\ (\citeyear{lu17}) have suggested that the
   CO excitation ladder of Mrk\,231 only significantly deviates from
   those of nearby starbursts like Arp\,220 and M82 in the \jco\ line
   and above.

\subsection{Implications for the cold gas density evolution}

\begin{figure}
  \vspace{-3mm}
\epsscale{1.28}
\plotone{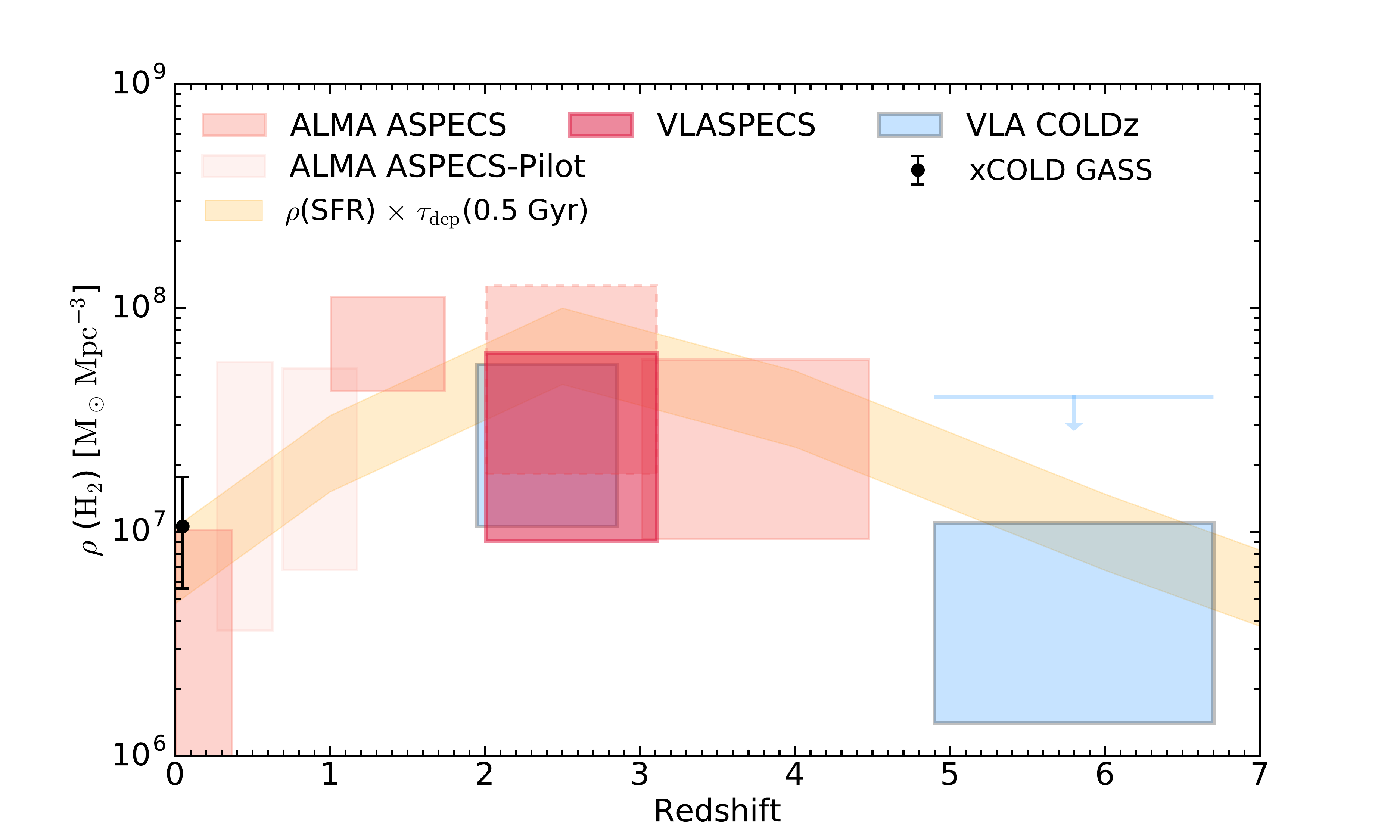}
\vspace{-5.5mm}

\caption{Constraints on the co-moving cold gas mass density evolution with redshift from the ASPECS (HUDF; salmon/light red boxes; Decarli et al.\ \citeyear{decarli16a,decarli19}) and COLDz surveys (COSMOS and GOODS-North combined; blue; Riechers et al.\ \citeyear{riechers19a}), and impact of the new VLASPECS measurements on the $z$$\sim$2--3 constraints from ASPECS (crimson red; corrected using the median $r_{31}$). Vertical sizes indicate uncertainties in each bin (2$\sigma$ for ASPECS; 90\% confidence region for COLDz). COLDz measurements are based on \aco\ at $z$=2.0--2.8 and \bco\ at $z$=4.9--6.7, whereas ASPECS measurements are based on \bco\ to \dco\ in the $z$$>$0.2 bins (including CO $J$=3$\to$2 at $z$=2.0--3.1), and \aco\ in the $z$$\sim$0 bin. 
    Other ASPECS redshift bins are left unscaled since no new constraints are available, but at least the $z$=0.3--0.6, 0.7--1.2, and 3.0--4.5 bins may also require a significant revision.
  The measurement at $z$=0 from the xCOLD GASS \aco\ survey (updated from Saintonge et al.\ \citeyear{saintonge17}) is shown for comparison. For reference, we also show the total star formation rate density multiplied by an equivalent gas depletion timescale of 0.5\,Gyr (Bouwens et al.\ \citeyear{bouwens16}). \label{f3}}
%\vspace{-5mm}
%
\end{figure}

Based on the ASPECS 3\,mm data and adopting $r_{31}$=0.42$\pm$0.07,
Decarli et al.\ (\citeyear{decarli19}) found a co-moving cosmic
molecular gas density of log($\rho({\rm
  H_2})$/\msol\,Mpc$^{-3}$)=7.26--8.10 (2$\sigma$) in the HUDF for the
$z$=2.0--3.1 redshift range. Adopting our median
$r_{31}$=0.84$\pm$0.26 at face value as the best estimate would reduce
this measurement to log($\rho({\rm
  H_2})$/\msol\,Mpc$^{-3}$)=6.96--7.80 (2$\sigma$),\footnote{The
  formal 1$\sigma$ range is log($\rho({\rm
    H_2})$/\msol\,Mpc$^{-3}$)=7.20--7.66.} with an average of 7.44. In
comparison, results from the COLDz survey in the COSMOS and
GOODS-North fields at $z$=2.0--2.8 (Riechers et
al.\ \citeyear{riechers19a}; see Fig.~\ref{f3}) suggest log($\rho({\rm
  H_2})$/\msol\,Mpc$^{-3}$)=7.04--7.75 (90\% confidence boundary),
with an average of 7.43.\footnote{Given the focus of this work, we
  here restrict the comparison to results from blank-field CO surveys,
  and defer comparisons to results from other methods (e.g., Scoville
  et al.\ \citeyear{scoville17}; Liu et al.\ \citeyear{liu19}; Lenkic
  et al.\ \citeyear{lenkic20}) to a future publication (R.\ Decarli et
  al.\ 2020, in preparation), but we note that the results from
    these studies are broadly consistent with those presented here.}
Thus, the constraints from both surveys in this redshift bin are
indistinguishable when adopting our new constraints on $r_{31}$.

The ASPECS constraints in the $z$=0.3--0.6 redshift interval are also
based on \cco\ measurements, whereas those at $z$=0.7--1.2, and
3.0--4.5 are based on \dco\ measurements, and they are scaled to line
ratios for the same reference sample as the $z$=2.0--3.1 bin (see
Decarli et al.\ \citeyear{decarli19}). Our new measurements suggest
that significant corrections may also be
required for those measurements. The remaining bins are based on
\aco\ and \bco\ measurements. Thus, the lowest-redshift bin at
$z$=0.0--0.4 is likely not affected by our new findings, while we
estimate that the $z$=1.0--1.7 bin is potentially affected at the
$\lesssim$10\%--20\% level. If confirmed, this would suggest a lower
redshift for the peak in the comoving gas density than previously
assumed.\footnote{These findings assume that the $\alpha_{\rm CO}$
  conversion factor for the galaxy populations dominating the signal
  does not change significantly with redshift, which is consistent
  with our current constraints.} In light of these findings, an
upcoming publication will quantitatively address the required changes
based on the full CO excitation ladders of all ASPECS galaxies in more
detail (B20), to fully assess the consequences of our new findings on
the cold gas density history of the universe.

\section{Summary and Conclusions} \label{sec:conclusions}

Using the VLA, we have measured \aco-based gas masses, gas depletion
times, and baryonic gas fractions for six galaxies discovered by the
ASPECS survey in the HUDF, and we obtained an upper limit for a
seventh source.\footnote{Since $\sim$48\% of the allocated time for
  the program remained unobserved from the present effort, three of
  the detections currently remain tentative.} This independently
confirms that these galaxies are gas-rich, and in some cases,
gas-dominated massive galaxies that are representative of the
``typical'' galaxy population at $z$=2--3 in terms of their star
formation rates (SFRs) and stellar masses.  Based on these
measurements, we revise previous estimates of the gas masses in this
redshift bin down by a factor of two on average. These findings
improve the agreement between measurements of the cold gas mass
density evolution with redshift from the ASPECS and COLDz surveys,
further demonstrating the reliability of the constraints obtained from
millimeter-wave line scan surveys across large cosmic
volumes. Comparing the ASPECS and COLDz samples (D.\ Riechers et
al.\ 2020, in preparation), there may be a hint that \cco\ selected
galaxies could have higher CO line excitation on average than
\aco\ selected galaxies, but current sample sizes are too small to
provide a firm conclusion.

The ASPECS ALMA survey was essential to identify these sources, which
would have been challenging with the VLA data alone. At the same time,
the longer-wavelength measurements carried out with the VLA are key to
extracting the most reliable constraints on the total gas masses and
the scales of any low-excitation gas reservoirs. In the near term
future, ALMA will be able to make similar measurements at $z$=1.2--2.3
with the addition of Band 1. Our findings suggest that future
facilities like the Next Generation Very Large Array (ngVLA; see,
e.g., Bolatto et al.\ \citeyear{bolatto17}) will only achieve their
full survey potential when including capabilities at both 9\,mm and
3\,mm, as is envisioned in the current baseline plan.

\acknowledgments

We thank the anonymous referee for a thorough and constructive
  report. D.R.\ acknowledges support from the National Science
Foundation under grant numbers AST-1614213 and AST-1910107. D.R. also
acknowledges support from the Alexander von Humboldt Foundation
through a Humboldt Research Fellowship for Experienced
Researchers. F.W.\ acknowledges support from the ERC Advanced grant
``Cosmic Gas''. I.R.S.\ acknowledges support from STFC
(ST/P000541/1). T.D-S. acknowledges support from the CASSACA and
  CONICYT fund CAS-CONICYT Call 2018. J.H. acknowledges support of
the VIDI research program with project number 639.042.611, which is
(partly) financed by the Netherlands Organization for Scientific
Research (NWO). H.I.\ acknowledges support from JSPS KAKENHI Grant
Number JP19K23462. M.K.\ acknowledges support from the International
Max Planck Research School for Astronomy and Cosmic Physics at
Heidelberg University (IMPRS-HD). Este trabajo cont\'o con el
  apoyo de CONICYT + PCI + INSTITUTO MAX PLANCK DE ASTRONOMIA
  MPG190030. The National Radio Astronomy Observatory is a facility
of the National Science Foundation operated under cooperative
agreement by Associated Universities, Inc. ALMA is a partnership
  of ESO (representing its member states), NSF (USA) and NINS (Japan),
  together with NRC (Canada), NSC and ASIAA (Taiwan), and KASI
  (Republic of Korea), in cooperation with the Republic of Chile. The
  Joint ALMA Observatory is operated by ESO, AUI/NRAO and NAOJ.

\appendix

\section{Upper limit spectrum for 9mm.7}

\begin{figure}[tbh]
\epsscale{0.6}
\plotone{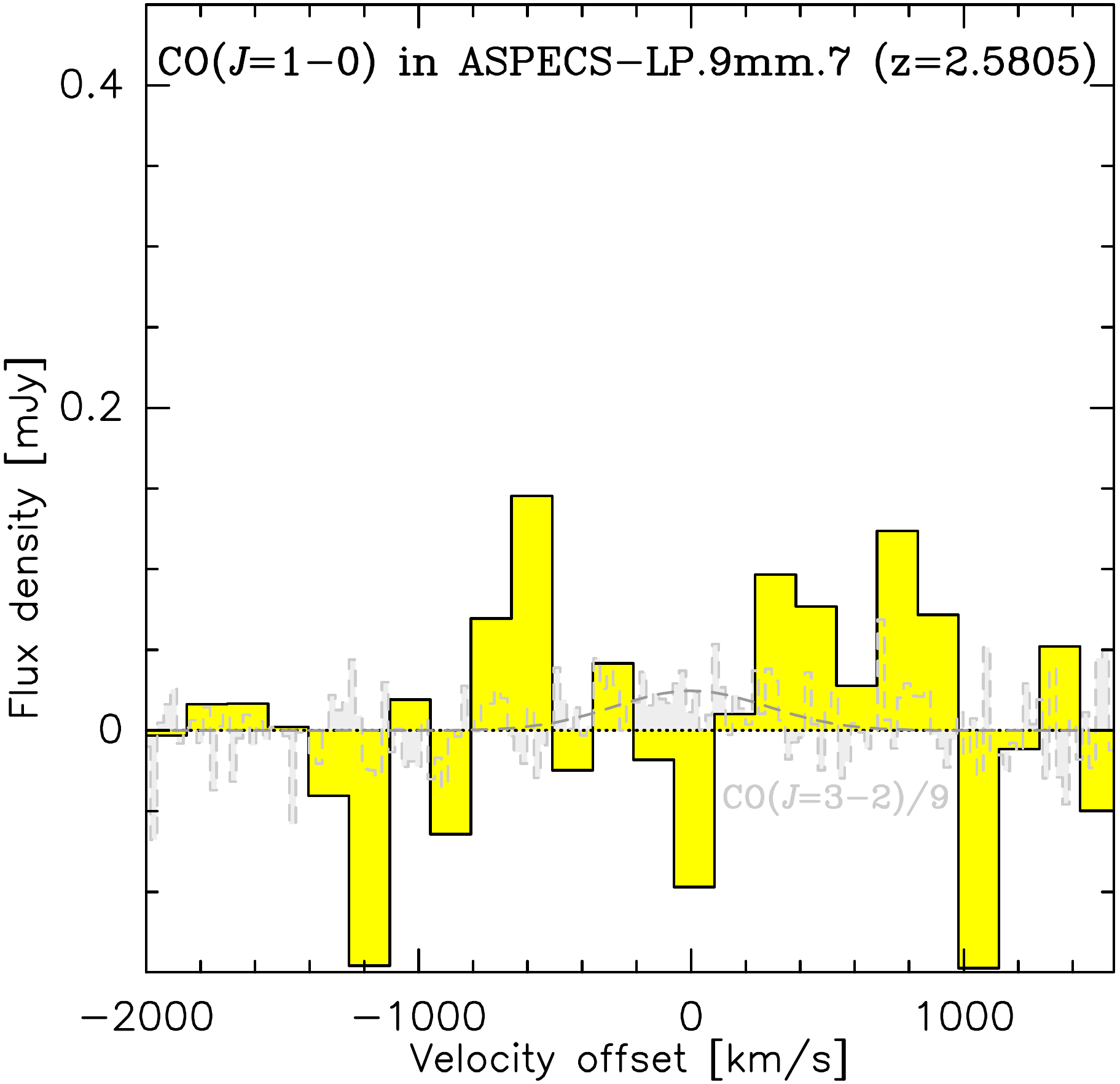}
\vspace{-2.0mm}

\caption{VLA upper limit \aco\ spectrum of 9mm.7 at a resolution of
  125\,\kms\ (16\,MHz), using the same style as in Fig.~\ref{f1}. \label{f2x}}
%\vspace{-5mm}

\end{figure}

The upper limit spectrum for 9mm.7 is shown in Fig.~\ref{f2x}. The
source is in a part of the mosaic with low primary beam response (see
Tab.~\ref{t1}), such that the VLA data are only moderately
constraining.

\facilities{VLA data:\ 19B-131, ALMA data:\ 2016.1.00324.L}

\bibliographystyle{yahapj}
\bibliography{ref.bib}

\end{document}